\documentclass[12pt]{article}%
\usepackage{amsmath,latexsym}
\usepackage{graphicx}
\usepackage{amsmath}
\usepackage{amsfonts}
\usepackage{amssymb}%
\begin{document}

\title{{A note on the predictability of flat galactic
    rotation curves}}
   \author{
Peter K. F. Kuhfittig*\\
\footnote{E-mail: kuhfitti@msoe.edu}
 \small Department of Mathematics, Milwaukee School of
Engineering,\\
\small Milwaukee, Wisconsin 53202-3109, USA}

\date{}
 \maketitle

\begin{abstract}\noindent
Based on an exact solution of the Einstein field
equations, it is proposed in this note that the
dark-matter hypothesis could have led to the
prediction of flat galactic rotation curves long
before the discovery thereof by assuming that on
large scales the matter in the Universe, including
dark matter, is a perfect fluid. \\

\end{abstract}

\section{Introduction}
\noindent
It has been known since the 1930's that galaxies and clusters of
galaxies do not have enough visible matter to account for
observed motions.  The missing matter was subsequently called
``dark matter."  The full implications of the existence thereof
was not recognized until the 1970's, when it was observed that
galaxies show solid-body rotations near the center but exhibit
flat rotation curves sufficiently far from the galactic core.
This indicates that the mass continues to increase linearly
with the radius, attributable to the presence of dark matter.
It is proposed in this note that flat rotation curves could
have been predicted from Einstein's theory prior to 1970 by
assuming that on large scales the matter in the Universe is
a perfect fluid that also includes the hypothesized dark
matter.  So we take the equation of state to be
$p=\omega\rho$ and then make use of the most general possible
exact solution of the Einstein field equations.

\section{Predictions from exact solutions}\noindent
We start with the line element
\begin{equation}\label{E:line1}
ds^{2}=-e^{2\Phi(r)}dt^{2}+e^{2\Lambda(r)}dr^2
+r^{2}(d\theta^{2}+\text{sin}^{2}\theta\,
d\phi^{2}).
\end{equation}
using units in which $c=G=1$.  The Einstein
field equations are
\begin{equation}\label{E:Einstein1}
   e^{-2\Lambda}\left(\frac{2\Lambda'}{r}-\frac{1}{r^2}
   \right)+\frac{1}{r^2}=8\pi\rho,
\end{equation}
\begin{equation}\label{E:Einstein2}
   e^{-2\Lambda}\left(\frac{1}{r^2}+\frac{2\Phi'}{r}
   \right)-\frac{1}{r^2}=8\pi p,
\end{equation}
and
\begin{equation}\label{E:Einstein3}
   e^{-2\Lambda}\left[(\Phi')^2+\Phi''-\Lambda'
   \Phi' +\frac{1}{r}\left(\Phi'-\Lambda'\right)
   \right]=8\pi p_t.
\end{equation}
Since Eq. (\ref{E:Einstein3}) can be obtained
from the conservation of the stress-energy tensor,
$T^{\mu\nu}_{\phantom{\mu\nu};\nu}=0$, only Eqs.
(\ref{E:Einstein1}) and (\ref{E:Einstein2}) are
needed.

Substituting in the equation of state $p=\omega\rho$,
we obtain the following differential equation:
\begin{equation}\label{E:diffequ}
   -\omega\Lambda'=-\Phi'+\frac{1}{2r}
   \left(e^{2\Lambda}-1\right)(\omega+1).
\end{equation}
This equation can be solved by separation of variables
to produce the most general possible exact solution if, and
only if, either $\Phi'\equiv 0$ ($\Phi\equiv\text{constant})$
or $\Phi$ is defined by $e^{2\Phi}=B_0r^l$, i.e., $\Phi'
=l/(2r)$, where $B_0$ is an arbitrary constant and $l$ is
fixed, first considered in Ref. \cite{pK16}.

In the former case ($\Phi'\equiv 0$), we obtain $e^{2\Lambda}
=1/(1-cr^{-(\omega +1)/\omega}$.  Now if $\omega =-1/3$, then
\begin{equation}
   e^{2\Lambda}=\frac{1}{1-cr^2},
\end{equation}
which leads to a special case of the
Friedmann-Lema\^{i}tre-Robertson-Walker (FLRW) model.
This outcome is consistent with the Friedmann
equation
\begin{equation}
   \frac{a''(t)}{a(t)}=-\frac{4\pi}{3}(\rho +3p)=
   -\frac{4\pi}{3}(\rho +3\omega\rho),
\end{equation}
yielding $a''(t)=0$ (no expansion) only if $\omega =-1/3$.
Letting $k^2=e^{2\Phi}$ and $c=K$, we obtain, after dividing
by $k^2$ and rescaling, the FLRW line element with
$a(t)\equiv 1$:
\begin{equation}\label{E:line2}
ds^{2}=-dt^{2}+\frac{dr^2}{1-Kr^2}
+r^{2}(d\theta^{2}+\text{sin}^{2}\theta\,
d\phi^{2}).
\end{equation}

In the present situation we are more interested in the case
$e^{2\Phi}=B_0r^l$ for $l>0$.  By Ref. \cite{pK16},
\begin{equation}\label{E:line3}
ds^{2}=-B_0r^ldt^{2}+\frac{dr^2}{\frac{\omega +1}{\omega +1+l}
   +Cr^{-(\omega +1+l)/\omega}}
+r^{2}(d\theta^{2}+\text{sin}^{2}\theta\,
d\phi^{2}),
\end{equation}
where $C$ is a constant of integration.  Keeping in mind
the line element \cite{MTW}
\begin{equation}\label{E:line4}
ds^{2}=-e^{2\Phi(r)}dt^{2}+\frac{dr^2}
   {1-\frac{2m(r)}{r}}
+r^{2}(d\theta^{2}+\text{sin}^{2}\theta\,
d\phi^{2}),
\end{equation}
where $m(r)$ is the total mass inside a sphere of
radius $r$, we rewrite Eq. (\ref{E:line3}) as
follows:
\begin{equation}\label{E:line5}
ds^{2}=-B_0r^ldt^{2}\\+\frac{dr^2}
   {1-2r(\frac{1}{2})[1-\frac{\omega +1}{\omega +1+l}
   -Cr^{-(\omega +1+l)/\omega}]/r}
+r^{2}(d\theta^{2}+\text{sin}^{2}\theta\,
d\phi^{2}).
\end{equation}
We now see that
\begin{equation}\label{E:mass}
     m(r)=r\left[\frac{1}{2}\left(1-\frac{\omega +1}
     {\omega +1+l}-Cr^{-(\omega +1+l)/\omega}
     \right)\right],
\end{equation}
where $0<\omega <1$ (ordinary or dark matter).
Since we are dealing with a region that is far
removed from the galactic center, the last term
in Eq. (\ref{E:mass}) becomes negligible.
So $m(r)$ has the linear form $m(r)=ar$:
\begin{equation}\label{E:m(r)}
  m(r)\approx\frac{1}{2}r\left(1-
  \frac{\omega +1}{\omega +1+l}\right).
\end{equation}
As already noted, this form implies the existence
of flat rotation curves and hence of dark matter.

Returning to line element (\ref{E:line3}), recall that
the exponent $l$ has to be a constant in the exact
solution obtained.  So $l$ has a natural
interpretation thanks to the well-established
model $B_0r^l$, where $l=2(v^{\phi})^2$ and
$v^{\phi}$ is the tangential velocity.  This
model is obtained from the Lagrangian
\begin{equation}
   \mathcal{L}=\frac{1}{2}\left(-e^{2\Phi(r)}
   \dot{t}^2+e^{2\Lambda(r)}\dot{r}^2
   +r^2(\dot{\theta}^2+\text{sin}^2\theta
   \,\dot{\phi}^2\right)
\end{equation}
by assuming a constant tangential velocity.
(See, for example, Ref. \cite{fR12}.)

With the benefit of hindsight this qualitative result proves to be
consistent with observation.  Suppose $m_1$ is the mass of a star,
$v^{\phi}$ its tangential velocity, and $m_2$ the total mass of
everything else.  (Here we need to use $m_2$ since $m(r)$ in
Eq. (\ref{E:m(r)}) is only an approximation.)  Then multiplying
$m_1$ by the centripetal acceleration yields
\begin{equation}\label{E:TF}
   m_1\frac{(v^{\phi})^2}{r}=m_1m_2\frac{G}{r^2}.
\end{equation}
Since $G=1$ and $(v^{\phi})^2=\frac{1}{2}l$, we obtain $m_2=
\frac{1}{2}lr$.  According to Ref. \cite{NSS}, $l\approx 0.000001$.
Since $m(r)<m_2$, Eq. (\ref{E:m(r)}) now leads to the requirement
$\omega >-0.000001$, which is met since $0<\omega <1$.

An example to illustrate Eq. (\ref{E:TF}) can be constructed
from Table 1 of Ref. \cite {NFW} on the Navarro-Frenk-White
model. The entries correspond to a ``virial" radius ranging
from 177 kpc for a dwarf galaxy to 3740 kpc for a rich galactic
cluster.  By choosing the large value of 342 kpc for $r$ in
Line 6, we may assume that the corresponding mass of
$2.301\times 10^{12}M_{\odot}$ corresponds to $m_2$ above.
The given value of $v^{\phi}=170.1$ km/s in the table agrees
with the value computed from Eq. (\ref{E:TF}) (with $G=
6.67\times 10^{-11}\, \text{N}\cdot\text{m}^2/\text{kg}^2$)
to three significant figures.

\section{Summary}

By starting with the most general possible exact solution
of the Einstein field equations given the perfect-fluid 
equation of state $p=\omega\rho$, it is proposed in this 
note that the following model for flat galactic rotation 
curves could have been predicted prior to 1970:
\begin{equation*}
ds^{2}=-B_0r^ldt^{2}+\frac{dr^2}
   {1-\frac{2m(r)}{r}}
+r^{2}(d\theta^{2}+\text{sin}^{2}\theta\,
d\phi^{2}),
\end{equation*}
where
\begin{equation*}
  m(r)\approx\frac{1}{2}r\left(1-
  \frac{\omega +1}{\omega +1+l}\right),
  \quad 0<\omega <1,
\end{equation*}
and $l$ is related to the tangential velocity.


\begin{thebibliography}{30}
\bibitem{pK16}P. K. F. Kuhfittig, Exactly solvable wormhole
   and cosmological models with a barotropic equation of
   state, Acta Physica Polonica B, 47, Issue 5, acepted,
   arXiv: 1408.4686.
\bibitem{MTW}C. W. Misner, K. S. Thorne and J. A. Wheeler,
   Gravitation (New York: W. Freeman and Company, 1973), page 608.
\bibitem{fR12}F. Rahaman, P. K. F. Kuhfittig, R. Amin,
   G. Mandal, S. Ray and N. Islam, Quark matter as dark matter
   in modeling galactic halo, Physics Letters B 714, Issues 2-5
   (2012), 131-135.
\bibitem{NSS}U. Nucamendi, M. Salgado and D. Sudarsky, An
   alternative approach to the galactic dark matter problem,
   Physical Review D 63 (2001), 125016.
\bibitem{NFW}J. F. Navarro, C. S. Frenk and S. D. M. White,
   The structure of cold dark matter, Astrophysical Journal
   462 (1996), 563-575.

\end{thebibliography}
\end{document}